\documentclass[twocolumn,english,prd,superscriptaddress,aps,showkeys,amsmath,amssymb,floatfix,nofootinbib]{revtex4-1}
\pdfoutput=1
\usepackage[pdftex]{graphicx}
\usepackage[pdftex]{color}
\usepackage[colorlinks,bookmarks]{hyperref}
\definecolor{linkblue}{rgb}{0,0,0.8}
\definecolor{linkgreen}{rgb}{0,0.5,0}
\definecolor{darkgreen}{rgb}{0,0.4,0}
\definecolor{purple}{rgb}{0.7,0.0,0.4}
\hypersetup{linkcolor=linkblue, citecolor=purple, urlcolor=linkblue}

\usepackage{nicefrac}

\usepackage[T1]{fontenc}
\usepackage[utf8]{inputenc}
\usepackage{lmodern}
\usepackage{babel}
\usepackage{bm}
\usepackage{verbatim}
\usepackage[us]{datetime}
\usepackage{enumerate}

\usepackage{latexsym}
\usepackage{mathrsfs}

\definecolor{vale}{rgb}{0,0.5, 1.}

\def\dd{{\rm d}}
\def\d{{\rm d}}
\def\be{\begin{equation}}
\def\ee{\end{equation}}

\begin{document}

\title{Null tests of the standard model using the linear model formalism}

\author{Valerio Marra}

\affiliation{Núcleo Cosmo-ufes \& Departamento de Física, Universidade Federal do Espírito Santo, 29075-910, Vitória, ES, Brazil}

\author{Domenico Sapone}

\affiliation{Departamento de Física, FCFM, Universidad de Chile, Blanco Encalada 2008, Santiago, Chile}

\begin{abstract}
We test both the FLRW geometry and $\Lambda$CDM cosmology in a model independent way by reconstructing the Hubble function $H(z)$, the comoving distance $D(z)$ and the growth of structure $f\sigma_8(z)$ using the most recent data available.
We use the linear model formalism in order to optimally reconstruct the above cosmological functions, together with their derivatives and integrals.
We then evaluate four of the null tests available in literature that probe both background and perturbation assumptions.
%We then evaluate four of the null tests available in literature: $Om_{1}$ by Sahni et al., $Om_{2}$ by  Zunckel \& Clarkson, $Ok$ by Clarkson et al., and $ns$ by Nesseris \& Sapone.
For all the four tests we find agreement, within the errors, with the standard cosmological model.
\end{abstract}


\maketitle

%%%%%%%%%%%%%%%%%%%%%%%%%%%%%%%%%%%%
%%%%%%%%%%%%%%%%%%%%%%%%%%%%%%%%%%%%
\section{Introduction}\label{intro}
%%%%%%%%%%%%%%%%%%%%%%%%%%%%%%%%%%%%
%%%%%%%%%%%%%%%%%%%%%%%%%%%%%%%%%%%%

The late-time accelerated expansion of the Universe has led cosmologists to revise the theory according to which the cosmos evolves either by introducing a new form of matter called dark energy~\cite{Sapone:2010iz} or by modifying directly the laws of gravity~\cite{Tsujikawa:2010zza}. 
Within the framework of Friedmann-Lema\^itre-Robertson-Walker (FLRW) cosmologies, a phase of accelerated expansion can be produced by a simple cosmological constant $\Lambda$.
Although the above gives rise to severe coincidence and fine-tuning problems, observations seem in agreement with such an explanation~\cite{Betoule:2014frx,Ade:2015xua,Abbott:2017wau}.

In order to make progress and understand what is causing the universe to accelerate one has, generally speaking, two options.
One can assume that the $\Lambda$CDM model is correct and work out its consequences. If inconsistencies with data are found, then physics beyond the standard model is necessary.
The other complementary approach is to assume a specific model beyond $\Lambda$CDM and work out if it has advantages with respect to the standard paradigm.
The second approach may be more powerful and may trigger new ideas and methodologies.
The first approach is simpler so that it is possible to study and understand the model phenomenology in much greater details.
The methodology of this paper belongs to the first approach.
In particular, we will analyze null tests of the standard model.

Null or consistency test analyses do not aim at finding the parameters of the model in question. Rather, they aim at uncovering possible tensions in data, which could be due to unaccounted-for systematics or a failure of the model itself.
These tests are model independent in the sense that they use directly the data, and have the added advantage that, if violated, one knows which set of theoretical assumptions have to be reanalyzed.
Equivalently, null tests have the ability to extract information that one may miss if restricting to parameter estimation.
It is worth stressing that it is imperative to corroborate the underlying cosmological model in such a way in view of future experiments that will collect an enormous amount of data, spanning over a wide range of redshift, see, for example, DES~\cite{Abbott:2017wau}, eBOSS~\cite{Blanton:2017qot}, J-PAS~\cite{Benitez:2014ibt}, DESI~\cite{Aghamousa:2016zmz}, LSST~\cite{Abell:2009aa}, Euclid~\cite{Amendola:2016saw}, SKA~\cite{Carilli:2004nx}.

Here we consider four null tests that have been proposed during the past 10 years: the $Om_{1}$ diagnostic by Sahni et al.~\cite{Sahni:2008xx}, the $Om_{2}$ diagnostic by  Zunckel \& Clarkson~\cite{Zunckel:2008ti}, the $Ok$ diagnostic by Clarkson et al.~\cite{Clarkson:2007pz}, and the $ns$ diagnostic by Nesseris \& Sapone~\cite{Nesseris:2014mfa}.
In order to evaluate these tests, we reconstruct the Hubble function $H(z)$, the comoving distance $D(z)$ and the growth of structure $f\sigma_8(z)$ using the most recent data available.
We use the linear model formalism in order to optimally reconstruct the above cosmological functions.
This method is simple and powerful as one can obtain an exact statistical description of the reconstructed functions, including their derivatives and integrals. Furthermore, it is an analytical approach which is suitable to be used with large datasets.
Therefore, we propose this method as a possible alternative to methods previously used in the literature, to list a few: binning data and discrete derivatives, principal component analysis, genetic algorithms, Padé approximation, Gaussian processes, non-parametric smoothing, machine learning~\cite{Sapone:2014nna,Escamilla-Rivera:2015odt,Gonzalez:2017tcm,Cao:2017gfv,Wang:2017sjw,LHuillier:2017ani}.

This paper is organized as follows.
In Section~\ref{basics} we review basic equations and the notation adopted in this paper, and in Section~\ref{nulls} we briefly present the four null tests considered in this work.
We describe the data we use in Section~\ref{data} and detail our methodology in Section~\ref{method}. The results of Section~\ref{results} show that the standard cosmological model passes all the tests.
The busy reader can jump to Figure~\ref{fom}.
We conclude in Section~\ref{conclusions}.

%%%%%%%%%%%%%%%%%%%%%%%%%%%%%%%%%%%%
%%%%%%%%%%%%%%%%%%%%%%%%%%%%%%%%%%%%
\section{Basic equations} \label{basics}
%%%%%%%%%%%%%%%%%%%%%%%%%%%%%%%%%%%%
%%%%%%%%%%%%%%%%%%%%%%%%%%%%%%%%%%%%

Here we review the basic equations upon which all the null tests are built. The evolution of the dark energy component can be expressed in terms of its present energy density parameter $\Omega_{de_0}$ and its equation of state parameter $w(z) = p/\rho$, being $p$ and $\rho$ the pressure and energy density, respectively.
The subscript $0$ will denote the present-day value of the corresponding quantity.
The Hubble parameter is then:
\be
\frac{H^2(z)}{H_0^2} = \Omega_{m_0}(1+z)^3 +\Omega_{de_0}(1+z)^{3(1+\hat{w})} +\Omega_{k_0}(1+z)^2
\label{eq:hubble-param}
\ee
where
\be
\hat{w}(z) = \frac{1}{\log (1+z)}\int_{0}^{z}\frac{w(z')}{1+z'}\dd z'\,,
\ee
and $\Omega_{m_0}$ and $\Omega_{k_0}$ are the matter and curvature density parameters, respectively.
If we are dealing with the cosmological constant, then $w(z) = -1$ and $\hat{w}(z)=-1$ at all redshifts. 
Furthermore, the relation $\Omega_{m_0}+\Omega_{de_0}+\Omega_{k_0}=1$ has to be satisfied.

In a general FLRW model with curvature, the angular diameter distance can be written as:
\be
D_A(z) = \frac{c}{1+z}\frac{1}{H_0\sqrt{-\Omega_{k_{0}}}}\sin\left(\sqrt{-\Omega_{k_{0}}}\int_{0}^{z}\frac{\dd z'}{H(z')}\right) \,.
\label{eq:angular-distance}
\ee
The angular diameter distance is related to the luminosity distance $D_{L}$ and the dimensionless comoving distance $D$ by the relations:
\begin{align}
D_{L}(z)&=(1+z)^{2} D_{A}(z) \,, \\
D(z) &=\frac{H_{0}}{c}  (1+z) D_{A}(z) \label{addi} \,.
\end{align}
In the Universe matter clusters forming perturbations $\delta\rho(t,x)$ to the underlying background energy density~$\rho(t)$. The growth of matter perturbation in the $\Lambda$CDM model is given by (assuming homogeneity and isotropy)
\be
\delta''(z)+\left(\frac{5}{1+z}-\frac{H'(z)}{H(z)}\right)\delta'(z) -\frac{3}{2}\frac{\Omega_{m}(z)}{(1+z)^2}\delta(z)=0
\label{eq:pert-growth}
\ee
where a prime refers to the derivative with respect to $z$, and the time evolving matter energy density is 
\be
\Omega_m(z) = \frac{\Omega_{m_0}(1+z)^3}{H^2(z)/H_0^2} \,.
\ee

%%%%%%%%%%%%%%%%%%%%%%%%%%%%%%%%%%%%
%%%%%%%%%%%%%%%%%%%%%%%%%%%%%%%%%%%%
\section{Null tests of the standard model} \label{nulls}
%%%%%%%%%%%%%%%%%%%%%%%%%%%%%%%%%%%%
%%%%%%%%%%%%%%%%%%%%%%%%%%%%%%%%%%%%

In this section we list the null tests that we consider in this paper; we refer to the corresponding literature for further details. 

%%%%%%%%%%%%%%%%%%%%%%%%%%%%%%%%%%%%
\subsection{$Om$ diagnostic}\label{Om}
%%%%%%%%%%%%%%%%%%%%%%%%%%%%%%%%%%%%

The $Om$ diagnostic was introduced to test deviations with respect to the flat $\Lambda$CDM scenario.
Setting $w=-1$ and $\Omega_{k_{0}}=0$ in Eq.~\eqref{eq:hubble-param}, one can solve for $\Omega_{m_{0}}$ and obtain the following diagnostic~\cite{Sahni:2008xx}:
%was introduced in \cite{Zunckel:2008ti,Sahni:2008xx} (see also \cite{Shafieloo:2009hi}) and
\begin{equation} \label{om1}
Om_{1}(z) = \frac{H(z)^{2}/H^{2}_{0}-1}{(1+z)^{3}-1} \,.
\end{equation}
Equivalently, one can solve for $\Omega_{m_{0}}$ in Eq.~\eqref{eq:angular-distance} and obtain the alternative diagnostic~\cite{Zunckel:2008ti}:
\begin{equation} \label{om2}
Om_{2}(z) = \frac{1/D'(z)^{2}-1}{(1+z)^{3}-1} \,.
\end{equation}
Clearly, within the flat $\Lambda$CDM model one has:
\begin{equation} \label{omre}
Om_{1}(z)=Om_{2}(z)=\Omega_{m_{0}} \,,
\end{equation}
which has to be valid at any redshift.
Any violation of the above relation will falsify the flat $\Lambda$CDM model.
As discussed below, we will obtain the luminosity distance $D_L(z)$ (and so $D$) from supernova Ia data, and the Hubble function $H(z)$ from the cosmic chronometer data.

%%%%%%%%%%%%%%%%%%%%%%%%%%%%%%%%%%%%
\subsection{$Ok$ diagnostic}\label{oktest}
%%%%%%%%%%%%%%%%%%%%%%%%%%%%%%%%%%%%

The following constant-curvature test can falsify not only the flat $\Lambda$CDM model but all the FLRW models at once.
The $Ok$ diagnostic is defined according to \cite{Clarkson:2007pz} (see also~\cite{LHuillier:2016mtc}):
\begin{equation}\label{ok}
Ok(z)=\frac{D'(z)^{2} H(z)^{2}/H^{2}_{0} -1}{D(z)^2} \,.
\end{equation}
If the FLRW models are the correct background models, one has:
\begin{equation} \label{okre}
Ok(z) = \Omega_{k_{0}} \,.
\end{equation}
Any violation not caused by standard-model perturbations%
\footnote{Standard-model perturbations can produce an additional systematic error on $H_{0}$~\cite{Marra:2013rba} and on the dark-energy equation of state~\cite{Valkenburg:2013qwa,Marra:2012pj}.}
would have profound implications as the FLRW model is at the basis of almost any cosmological model (except inhomogeneous~\cite{Valkenburg:2012td} and backreaction models, see the CQG special issue \cite{Andersson:2011za}, and also~\cite{Montanari:2017yma}).

%%%%%%%%%%%%%%%%%%%%%%%%%%%%%%%%%%%%
\subsection{$ns$ diagnostic}\label{NStest}
%%%%%%%%%%%%%%%%%%%%%%%%%%%%%%%%%%%%

While the previous tests probe the background structure of the universe, the test proposed in~\cite{Nesseris:2014mfa} (and deeply investigated in \cite{Nesseris:2014qca}) is sensitive to both background and perturbation observables. 
The $ns$ diagnostic is able to test the validity of the growth of matter perturbations under the assumption of an FLRW metric with a cosmological constant (which has no perturbations). In brief, the $ns$ diagnostic is obtained from Eq.~\eqref{eq:pert-growth}: first, the corresponding Lagrangian is found, then, with the help of Noether's theorem, the associated conserved quantity is obtained.
The conserved quantity is the $ns$ diagnostic that is found to be: 
\begin{eqnarray} \label{ns}
&&ns(z)=\frac{1}{1+z}\frac{H(z)}{H(0)}\frac{f\sigma_8(z)}{f\sigma_8(0)}\times\nonumber \\
&&\exp\left[\frac{3\cdot10^4}{2}\omega_{m}\int_{0}^{z}(1+x)^2\frac{\sigma_{8,0}-\int_{0}^x\frac{f\sigma_8(y)}{1+y}dy}{H(x)^2f\sigma_8(x)}\dd x\right] \,
\end{eqnarray}
where $\sigma_{8,0}$ is the normalization of the power spectrum (present day mass fluctuation on a scale of $8 h^{-1}$Mpc, this refers to linear
perturbation theory), and:
\begin{equation}
f\sigma_8(z)= f(z) G(z) \sigma_{8,0} \,,
\end{equation}
in which $G(z)$ is the growth function of matter perturbations normalized to unity today ($G(z)=\delta(z)/\delta(0)$, from \eqref{eq:pert-growth}) and the growth factor is $f= \d \ln G/\d \ln (1+z)+1$.
%, where the growth factor is defined through the density contrast as $\delta(z) = G(z)/(1+z)$.

In order to be robust against systematics of a particular experiment, we need to reconstruct the $ns(z)$ test by using four independent observables: the Hubble parameter $H(z)$, the growth of structure $f\sigma_8(z)$, $\sigma_{8,0}$ and $\omega_m = \Omega_{m_0}h^2$.
As discussed below, we will obtain $f\sigma_8$ from the RSD measurements collected by different experiments; the Hubble data come from the cosmic chronometers; instead, for $\sigma_{8,0}$ we will use alternatively the results from the SDSS-III BOSS~\cite{Tojeiro:2012rp} and KiDS~\cite{Joudaki:2017zdt} surveys, and for $\omega_m$ the results from the Planck satellite~\cite{Ade:2015xua}. 

Within any (not necessarily flat) $\Lambda$CDM model one has:
\begin{equation} \label{nsre}
ns(z)=1 \,.
\end{equation}
Any violation of the above relation may imply a deviation from the FLRW model, nonzero dark-energy perturbations and/or a deviation from GR.

%%%%%%%%%%%%%%%%%%%%%%%%%%%%%%%%%%%%
%%%%%%%%%%%%%%%%%%%%%%%%%%%%%%%%%%%%
\section{Data} \label{data}
%%%%%%%%%%%%%%%%%%%%%%%%%%%%%%%%%%%%
%%%%%%%%%%%%%%%%%%%%%%%%%%%%%%%%%%%%

The four tests above use three cosmological functions: $H(z)$, $D(z)$ and $f\sigma_8(z)$. We will estimate these functions using three distinct datasets. We will use the full covariance matrix when available.

\subsection{Cosmic chronometers}\label{ccd}

\begin{figure}[h]
\begin{centering}
\includegraphics[width=\columnwidth]{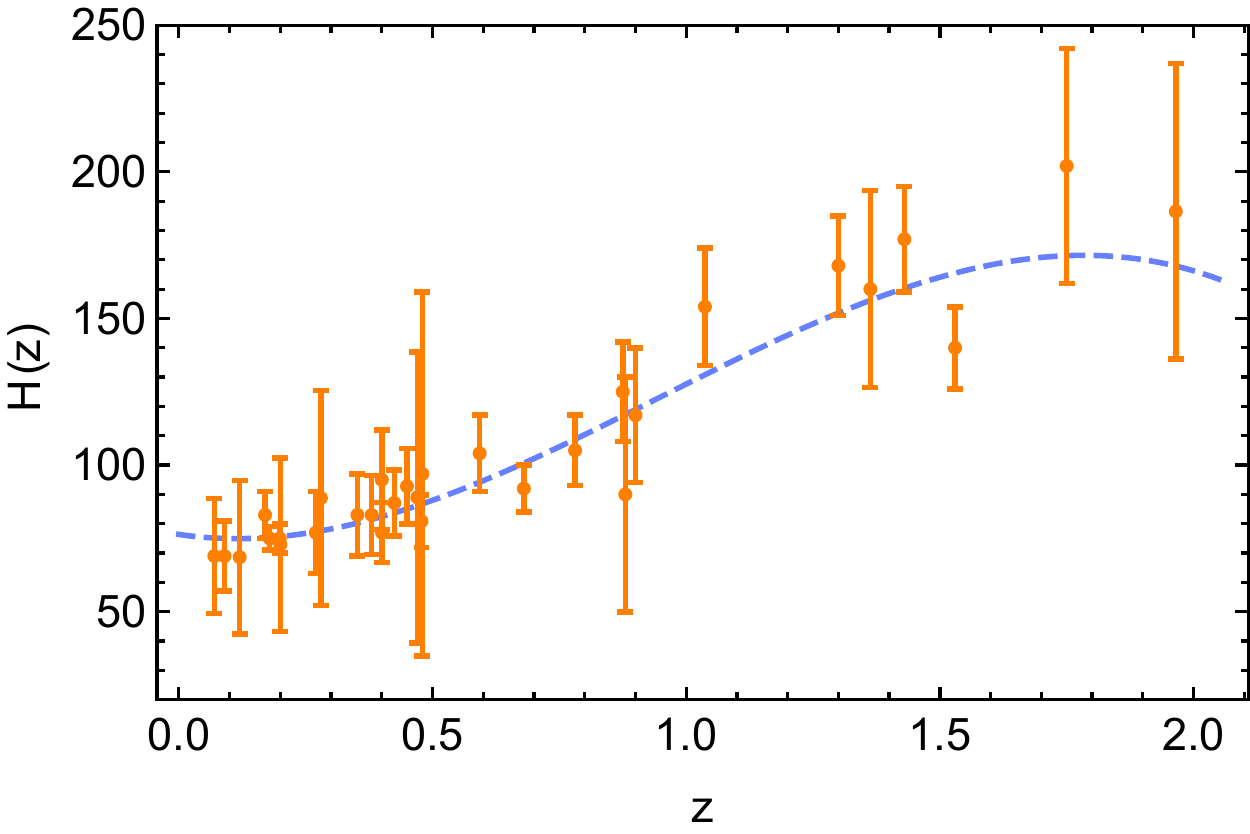}
\caption{The 31 cosmic chronometer data points and the corresponding linear model best fit (in light blue) that we use in this work. See Table~\ref{tab:tableH} for the numerical values.}
\label{ccdata}
\end{centering}
\end{figure}

The so-called ``cosmic chronometers'' are passively evolving galaxies from which it is possible to obtain in a model-independent way the Hubble parameter $H(z)$ at various redshifts~\cite{Jimenez:2001gg}.
Here, we use the 31 independent data points given in Table~\ref{tab:tableH}.
This is the most up-to-date collection of $H(z)$ data: see Figure~\ref{ccdata} for a plot (where also shown is the corresponding linear model best fit, see Section~\ref{method}).

\begin{table}[htp]
\caption{The 31 cosmic chronometer data points used in this analysis.}
\begin{center}
\begin{tabular}{ccccccccc}
\hline
\hline
$z$ & $H(z)$ & $\sigma_{H(z)}$& ref. & \phantom{jholmes}&$z$ & $H(z)$ & $\sigma_{H(z)}$ & ref.\\
\hline
 0.07 & 69.0 & 19.6 & \cite{Zhang:2012mp} & &0.4783 & 80.9 & 9.0  & \cite{Moresco:2016mzx} \\
 0.09 & 69.0 & 12.0 & \cite{Simon:2004tf} & & 0.48 & 97.0 & 62.0 & \cite{Stern:2009ep}  \\
 0.12 & 68.6 & 26.2 & \cite{Zhang:2012mp}  & & 0.593 & 104.0 & 13.0 & \cite{Moresco:2012jh} \\
 0.17 & 83.0 & 8.0 & \cite{Simon:2004tf}  & & 0.68 & 92.0 & 8.0 & \cite{Moresco:2012jh} \\
 0.179 & 75.0 & 4.0 & \cite{Moresco:2012jh}  & & 0.781 & 105.0 & 12.0  & \cite{Moresco:2012jh} \\
 0.199 & 75.0 & 5.0 & \cite{Moresco:2012jh}  & & 0.875 & 125.0 & 17.0 & \cite{Moresco:2012jh} \\
 0.2 & 72.9 & 29.6 & \cite{Zhang:2012mp}  & & 0.88 & 90.0 & 40.0 & \cite{Stern:2009ep} \\
 0.27 & 77.0 & 14.0 & \cite{Simon:2004tf}  & & 0.9 & 117.0 & 23.0 & \cite{Simon:2004tf} \\
 0.28 & 88.8 & 36.6 & \cite{Zhang:2012mp}  & & 1.037 & 154.0 & 20.0 & \cite{Moresco:2012jh} \\
 0.352 & 83.0 & 14.0 & \cite{Moresco:2012jh}  & & 1.3 & 168.0 & 17.0 & \cite{Simon:2004tf} \\
 0.3802 & 83.0 & 13.5  & \cite{Moresco:2016mzx} & & 1.363 & 160.0 & 33.6  & \cite{Moresco:2015cya} \\
 0.4 & 95.0 & 17.0 & \cite{Simon:2004tf}  & & 1.43 & 177.0 & 18.0 & \cite{Simon:2004tf} \\
 0.4004 & 77.0 & 10.2 & \cite{Moresco:2016mzx}  & & 1.53 & 140.0 & 14.0 & \cite{Simon:2004tf} \\
 0.4247 & 87.1 & 11.2  & \cite{Moresco:2016mzx} & & 1.75 & 202.0 & 40.0 & \cite{Simon:2004tf} \\
 0.4497 & 92.8 & 12.9  & \cite{Moresco:2016mzx} & & 1.965 & 186.5 & 50.4  & \cite{Moresco:2015cya} \\
 0.47 & 89.0 & 49.6  & \cite{Ratsimbazafy:2017vga} & & & & & \\
\hline
\hline
\end{tabular}
\end{center}
\label{tab:tableH}
\end{table}

\subsection{Supernovas Ia}\label{snia}

\begin{figure}[h]
\begin{centering}
\includegraphics[width=\columnwidth]{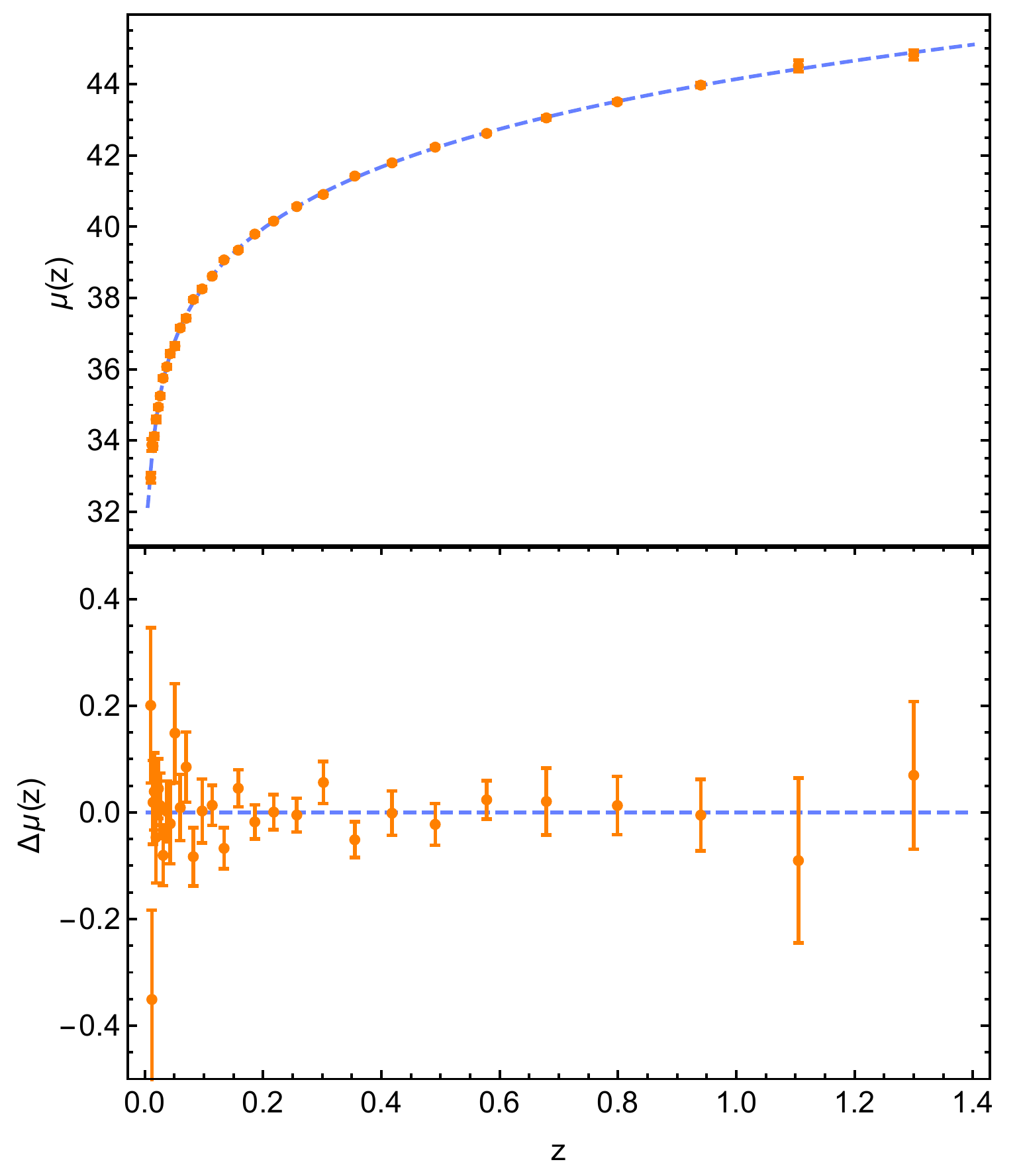}
\caption{The compressed JLA compilation \cite[][Appendix E]{Betoule:2014frx}  and the corresponding linear model best fit (in light blue) that we use in this work. For this plot we have adopted $M=0$.}
\label{JLA}
\end{centering}
\end{figure}

We obtain the luminosity distance (in Mpc) from the distance modulus $\mu$:
\begin{equation}
D_{L}(z)=10^{\frac{\mu(z)-25}{5}} \,.
\end{equation}
The distance modulus can be obtained from supernova Ia data. We will consider the compressed JLA compilation~\cite[][Appendix E]{Betoule:2014frx} so that:
\begin{equation}
\mu(z) = \mu_{b}(z)-M \,,
\end{equation}
where $\mu_{b}(z)$ is the binned distance modulus and $M$ is an unknown offset (a nuisance parameter). Fig.~\ref{JLA} shows the JLA data.

\subsection{Redshift space distortion data}\label{rsd}

\begin{figure}[h]
\begin{centering}
\includegraphics[width=\columnwidth]{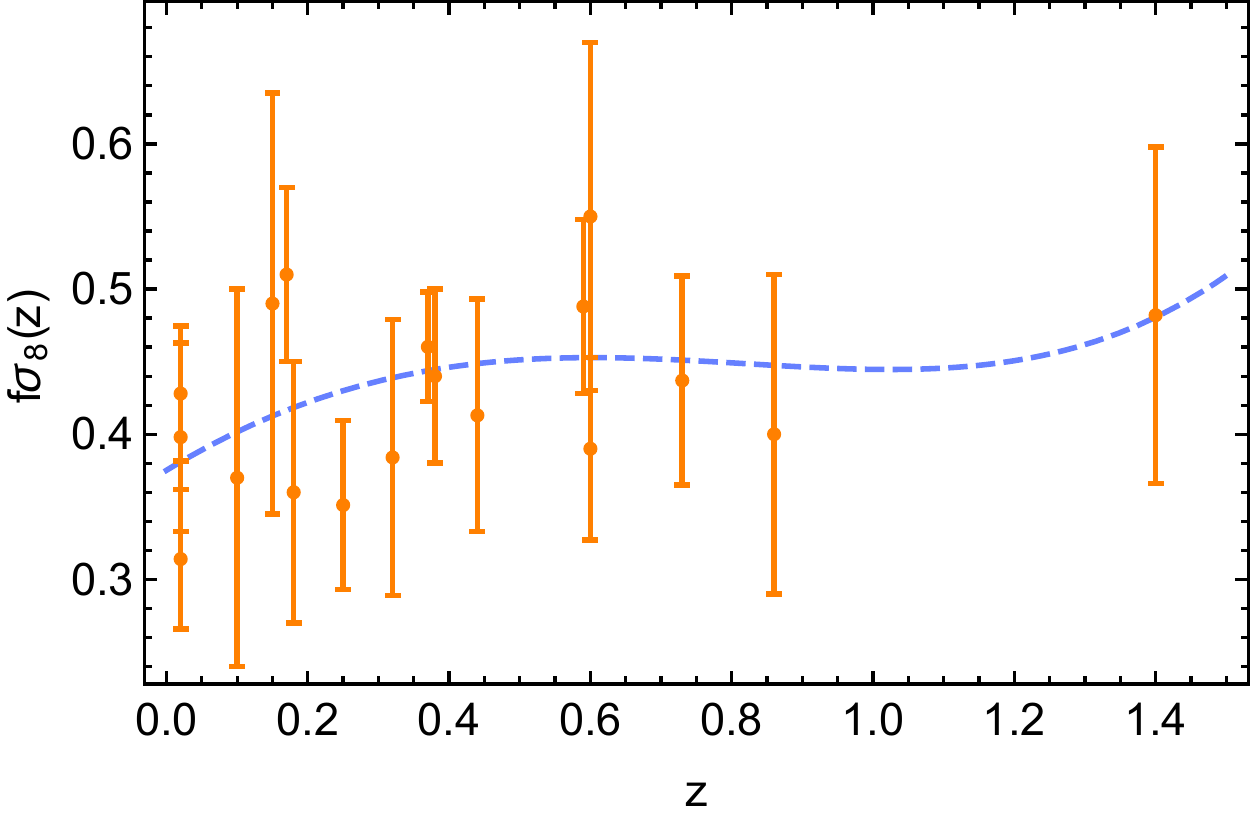}
\caption{The 18 $f\sigma_8$ data points~\cite{Nesseris:2017vor} and the corresponding linear model best fit (in light blue) that we use in this work.}
\label{fs8}
\end{centering}
\end{figure}

In order to reconstruct $f\sigma_8(z)$ we will consider the robust and independent measurements given in the ``Gold'' RSD compilation from \cite{Nesseris:2017vor}, see Fig.~\ref{fs8}.
 
\ 

The three sets of data used in this paper come from several surveys that span different angular and redshift ranges. Therefore, it should be justified to assume that we are considering a fair sample of the universe.

\subsection{Addition parameters} \label{additional}

For the parameter $\omega_m$, necessary for the $ns$ test, we use the results from Planck \cite[][Table 4, TT,TE,EE+lowP]{Ade:2015xua}:
\begin{align}
\omega_m  &= 0.1427 \pm 0.0014 \,.
\end{align}
For the parameter $\sigma_{8,0}$, also necessary for the $ns$ test, we use three different values from two surveys. Indeed, the Planck cosmological constraints on $\sigma_{8,0}$ are in
tension with those from Planck clusters~\cite{Ade:2015fva} and from weak lensing measurements, and it is important to test how  different values of $\sigma_{8,0}$ impact our results.
We will consider results from the SDSS-III BOSS~\cite{Tojeiro:2012rp} and KiDS~\cite{Joudaki:2017zdt} surveys, which are independent from Planck. SDSS is a spectroscopic galaxy survey while KiDS is a survey that uses cosmic shear, galaxy-galaxy lensing and redshift-space distortion galaxy clustering measurements. The values we consider are:
\begin{align} \label{s8va}
\sigma_{8,0} &= 0.804 \pm 0.051 \text{ SDSS-III BOSS} \,, \\
\sigma_{8,0} &= 0.832 \pm 0.080 \text{ KiDS} \,, \nonumber \\
\sigma_{8,0} &= 0.747 \pm 0.109 \text{ KiDS (conservative cut)} \nonumber \,.
\end{align}
The Hubble constant $H_{0}$ can be determined using the reconstructed $H(z)$ function at $z=0$.
However, it may also be interesting to consider an independent prior on $H_{0}$. This could be useful to test for systematics. We will consider results from local measurements~\cite{Riess:2016jrr} and from the CMB~\cite[][Table 4, TT,TE,EE+lowP]{Ade:2015xua}, respectively:
\begin{align}
H_{0,\rm{loc}}    &= 73.24 \pm 1.74 \; \frac{\text{km/s}}{\text{Mpc}} \,, \label{riess} \\
H_{0,\rm{cmb}}  &= 67.27 \pm 0.66 \; \frac{\text{km/s}}{\text{Mpc}} \,. \label{cmb}
\end{align}
The parameter $M$ enters in the determination of the dimensionless comoving distance $D$ that is used in the tests $Om_{2}$ and $Ok$. As $M$ is degenerate with $-5 \log_{10} H_{0}$, we leave $M$ free.

%%%%%%%%%%%%%%%%%%%%%%%%%%%%%%%%%%%%
\section{Method} \label{method}
%%%%%%%%%%%%%%%%%%%%%%%%%%%%%%%%%%%%

In order to reconstruct the cosmological functions, and also their derivatives and integrals, we will use the linear model formalism; see, for instance, \cite{gregory2005bayesian}.

%%%%%%%%%%%%%%%%%%%%%%%%%%%%%%%%%%%%
\subsection{Linear model analysis} \label{lma}
%%%%%%%%%%%%%%%%%%%%%%%%%%%%%%%%%%%%

Let us choose a set of base functions $g_{\alpha}(z)$ whose {\it linear} combination will constitute the template function $t(z,c_{\alpha})$:
\begin{equation} \label{tempe1}
t(z,c_{\alpha}) = \sum_{\alpha=0}^{\alpha_{\rm max}} c_\alpha \, g_{\alpha}(z) \,,
\end{equation}
where $\alpha$ is an integer.
The assumption is that $t(z,c_{\alpha})$ can describe the actual functions that we want to reconstruct: $H(z)$, $\mu(z)$ or $f\sigma_8(z)$. 
Clearly, this is conditional to an appropriate choice of $g_{\alpha}(z)$ and $\alpha_{\rm max}$, for each of the functions $H(z)$, $\mu(z)$ and $f\sigma_8(z)$.
We call $\alpha_{\rm max}$ the ``order'' of the template, which will then have $\alpha_{\rm max}+1$ coefficients.

Let us then assume that the data are given by:
\begin{equation}
d_i = t_i + e_i \,,
\end{equation}
where $t_i=t(z_i,c_{\alpha})$ and $e_i$ are Gaussian errors with covariance matrix $C_{ij}$.

Next we fit the template $t$ to the data and use the linear model formalism to calculate the Fisher matrix relative to the parameters $c_\alpha$.
This gives an exact description of the likelihood as the template is linear in its parameters.
The Fisher matrix is:
\begin{equation} \label{fisher}
F_{\alpha \beta} = g_{\beta i} C^{-1}_{i j} g_{\alpha j}
\end{equation}
where $g_{\alpha i}= g_{\alpha}(z_{i})$, and the best-fit values of $c_\alpha$ are:
\begin{equation}
c_{\alpha, \rm bf}= F^{-1}_{\alpha \beta} B_\beta \,,
\end{equation}
where $B_\alpha = d_i C^{-1}_{i j} g_{\alpha j}$.

Summarizing, we have propagated the covariance matrix $C_{ij}$ into the covariance matrix $F^{-1}_{\alpha \beta}$ on the parameters.

%%%%%%%%%%%%%%%%%%%%%%%%%%%%%%%%%%%%
\subsection{Error propagation} \label{error}
%%%%%%%%%%%%%%%%%%%%%%%%%%%%%%%%%%%%

Let us denote with $\phi(z,\theta_{\alpha})$ either $Om_{1}$, $Om_{2}$, $Ok$ or $ns$. $\phi(z,\theta_{\alpha})$ will be a nonlinear function of the various templates $t(z,c_{\alpha})$ (one for each $H(z)$, $\mu(z)$ and $f\sigma_8(z)$) and their derivatives and integrals.
The parameter vector~$\{\theta_{\alpha} \}$ comprises the template parameters of Section~\ref{lma} and the additional parameters of Section~\ref{additional} that enter $\phi(z,\theta_{\alpha})$.
The corresponding covariance matrix $\Sigma_{\alpha \beta}$ is obtained by forming an appropriate block diagonal matrix using the covariance matrices of the corresponding parameters.
We have chosen independent data (see Section~\ref{data}) so that correlations among different sets of data are not expected to be important.

In order to compute the error on $\phi(z,\theta_{\alpha})$ due to the uncertainty encoded in the covariance matrix $\Sigma_{\alpha \beta}$, a straightforward approach is to apply a change of variable from $\{\theta_{\alpha}\}$ to $\phi$.
At the first order, the error is then given by:
\begin{align} \label{linear}
\sigma^{2}_{\phi} = J_{\alpha}\Sigma_{\alpha \beta} J_{\beta} \,,
\end{align}
where
\begin{equation}
J_{\alpha} =\left. \frac{\partial \phi(z,\theta_{\alpha})}{\partial \theta_{\alpha}} \right|_{\theta_{\alpha, \rm bf}} \,.
\label{eq:jacobian-matrix}
\end{equation}
Eq.~\eqref{linear} is exact if $\phi(z,\theta_{\alpha})$ is $t(z,c_{\alpha})$, its derivative or integral. Indeed in this case it will depend linearly on the parameters $\{\theta_{\alpha}\}$.

%%%%%%%%%%%%%%%%%%%%%%%%%%%%%%%%%%%%
\subsection{Choice of base functions} \label{base}
%%%%%%%%%%%%%%%%%%%%%%%%%%%%%%%%%%%%

We will adopt the following base functions:
\begin{align}
H(z)  &\longrightarrow g_{\alpha}(z) = z^{\alpha} &\text{ with } 0\le \alpha \le \alpha_{\rm max}  \,, \nonumber \\
\mu(z)  & \longrightarrow g_{\beta}(z) = (\ln z)^{\beta} &\text{ with } 0\le \beta \le \beta_{\rm max} \,, \nonumber \\
f\sigma_8(z) & \longrightarrow g_{\gamma}(z) = (1+z)^{\gamma} &\text{ with } 0\le \gamma \le \gamma_{\rm max}\,. \nonumber
\end{align}
We have chosen the latter as they can reproduce the fiducial $\Lambda$CDM functions with not too high orders $\alpha_{\rm max}$, $\beta_{\rm max}$ and $\gamma_{\rm max}$, that is, with not too many parameters.
Furthermore, the base functions chosen for $f\sigma_8(z)$ allow us to compute analytically the inner integral in Eq.~\eqref{ns}.

In order to choose the orders $\alpha_{\rm max}$, $\beta_{\rm max}$ and $\gamma_{\rm max}$ on which the template functions $t(z,c_{\alpha})$ depend, see \eqref{tempe1}, we use the following strategy:
\begin{enumerate}

\item We pick values for $\{ \alpha_{\rm max}, \beta_{\rm max}, \gamma_{\rm max}\}$. Each order has to be within 0 and $N-1$, where $N$ is the number of data of the corresponding catalog.\footnote{With $N$ points one can fit up to $N$ coefficients so that, for example, the order $\alpha_{\rm max}$ cannot exceed $N-1$.}
\label{open}

\item We reconstruct $H(z)$, $\mu(z)$ and $f\sigma_8(z)$ using the method of Section~\ref{lma} with mock catalogs created using a fiducial $\Lambda$CDM cosmology.
These mocks share the same redshift values and covariance matrix of the real data but do not have fluctuations. In other words, the corresponding $H(z)$, $\mu(z)$ and $f\sigma_8(z)$ values are exactly the fiducial one.
Mock catalogs without fluctuations are often used when one wants to consider the average behavior of a large number of ``real life'' mocks with fluctuations.

\item We evaluate the four null tests of Section~\ref{nulls}.

\item We obtain the sigma bands of the null test functions using the results of Section~\ref{error}.

\item We calculate the following vector:
\begin{equation}
\Delta_{\phi}= \left\{ \frac{|\phi(z_{i},\theta_{\alpha, \rm bf})-\phi_{\rm fid}(z_{i})|}{\sigma_{\phi}(z_{i},\theta_{\alpha, \rm bf})} \right\} \,,
\end{equation}
with $i=1, \dots, N$.
\label{close}

\item
We set a reconstruction accuracy $Q$ and we repeat the steps \ref{open}-\ref{close} until we find the smallest values $\{ \alpha_{\rm max}, \beta_{\rm max}, \gamma_{\rm max}\}$ so that the following  condition is satisfied:
\begin{align}
\max \{ \Delta_{Om_{1}},\Delta_{Om_{2}},\Delta_{Ok},\Delta_{ns} \} < Q \,.
\end{align}

\end{enumerate}
By setting $Q=0.1$, we find:
\begin{equation}
\{ \alpha_{\rm max}, \beta_{\rm max}, \gamma_{\rm max}\}= \{ 3, 6,4\} \,.
\end{equation}
Figures \ref{ccdata}-\ref{fs8} show the reconstruction of the cosmological functions that we obtained.

The above strategy guarantees that the template functions can replicate the fiducial $\Lambda$CDM functions without inserting modeling biases in the analysis while, at the same time, keeping the template order as low as possible.
It is worth stressing that within our methodology a unnecessarily high order would lead to higher uncertainties in the Fisher matrix of equation~\eqref{fisher} thus degrading the constraining power of the null tests.

%%%%%%%%%%%%%%%%%%%%%%%%%%%%%%%%%%%%
%%%%%%%%%%%%%%%%%%%%%%%%%%%%%%%%%%%%
\section{Results} \label{results}
%%%%%%%%%%%%%%%%%%%%%%%%%%%%%%%%%%%%
%%%%%%%%%%%%%%%%%%%%%%%%%%%%%%%%%%%%

\begin{figure*}
\begin{centering}
\includegraphics[width=\columnwidth]{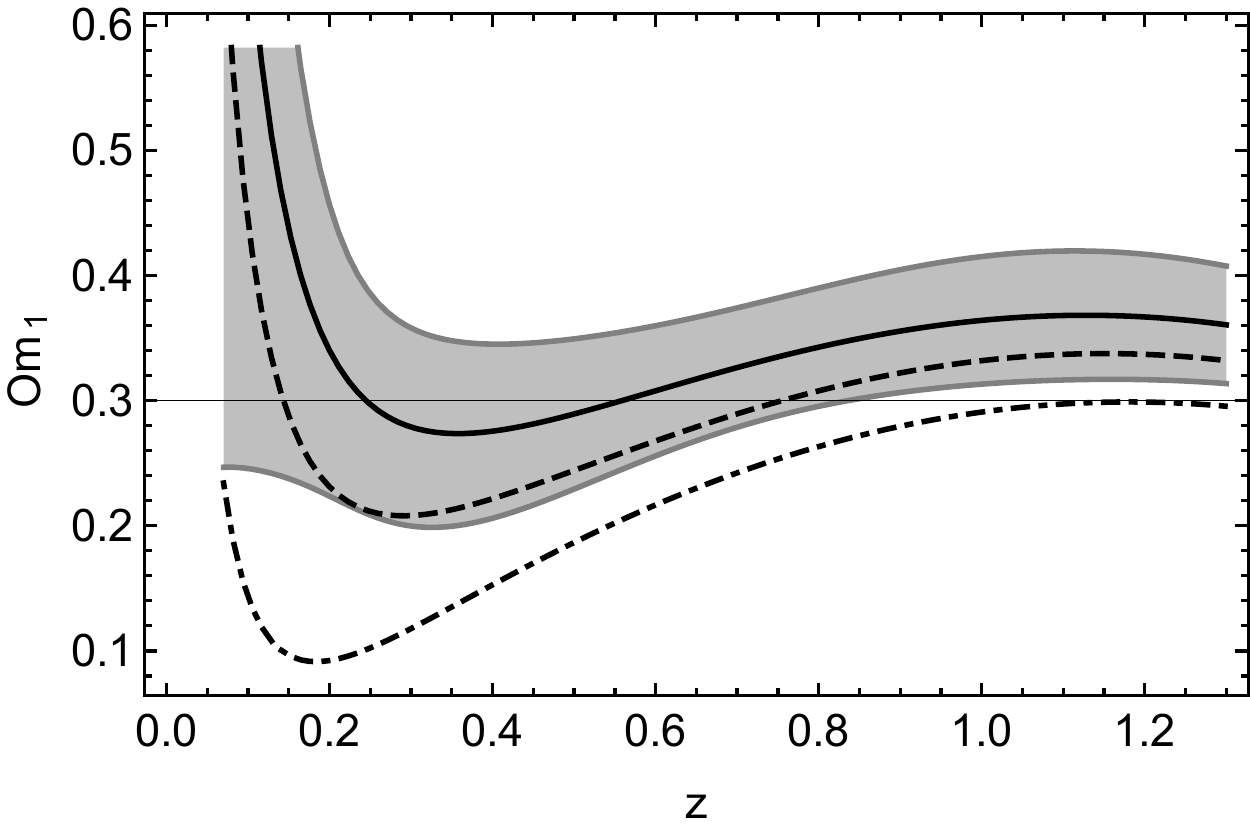}
\includegraphics[width=\columnwidth]{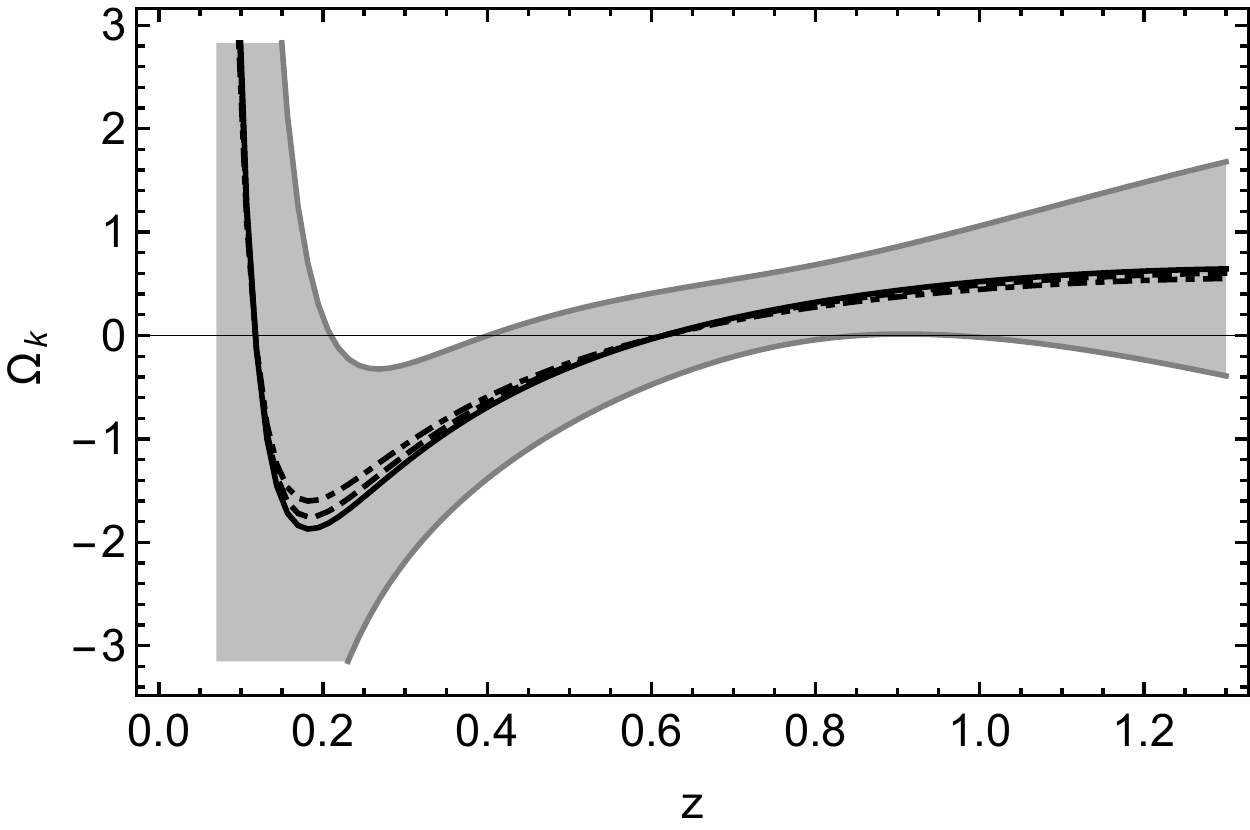}\\
\vspace{.4cm}
\includegraphics[width=\columnwidth]{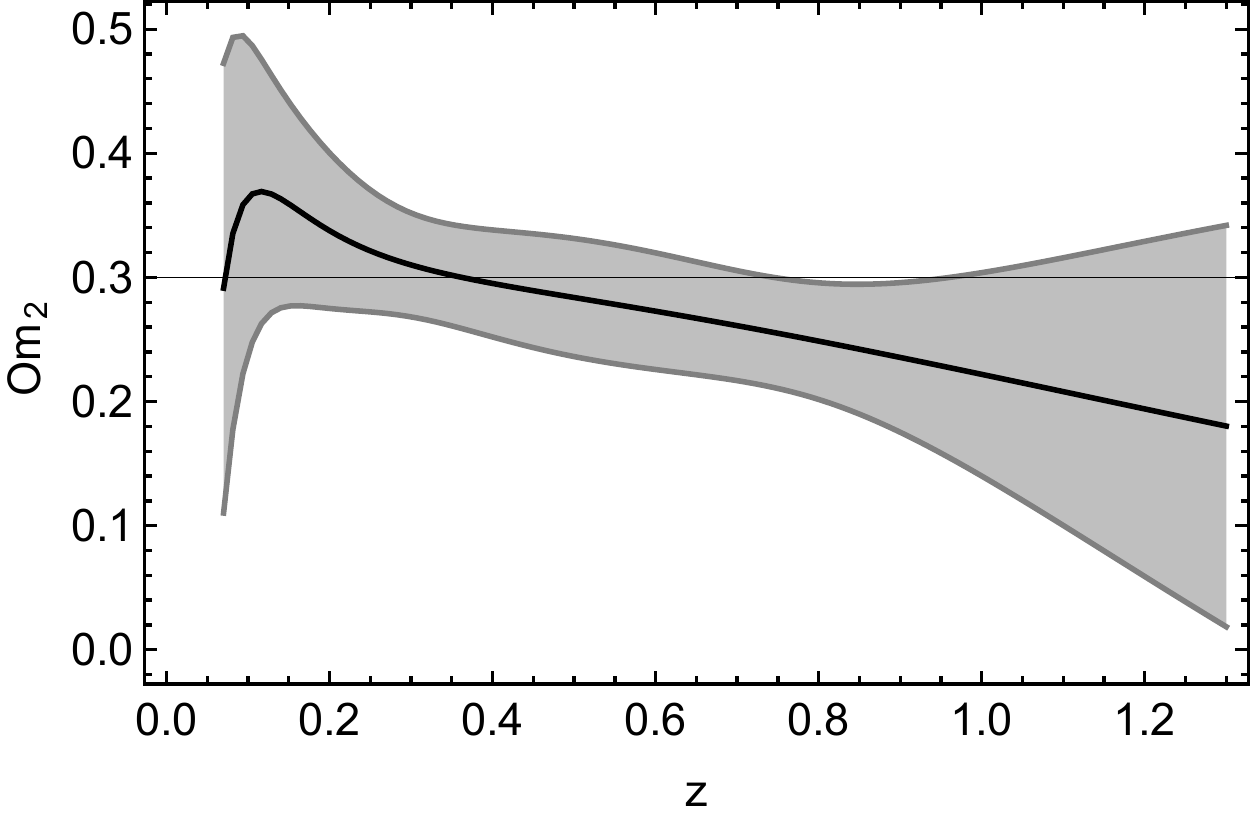}
\includegraphics[width=\columnwidth]{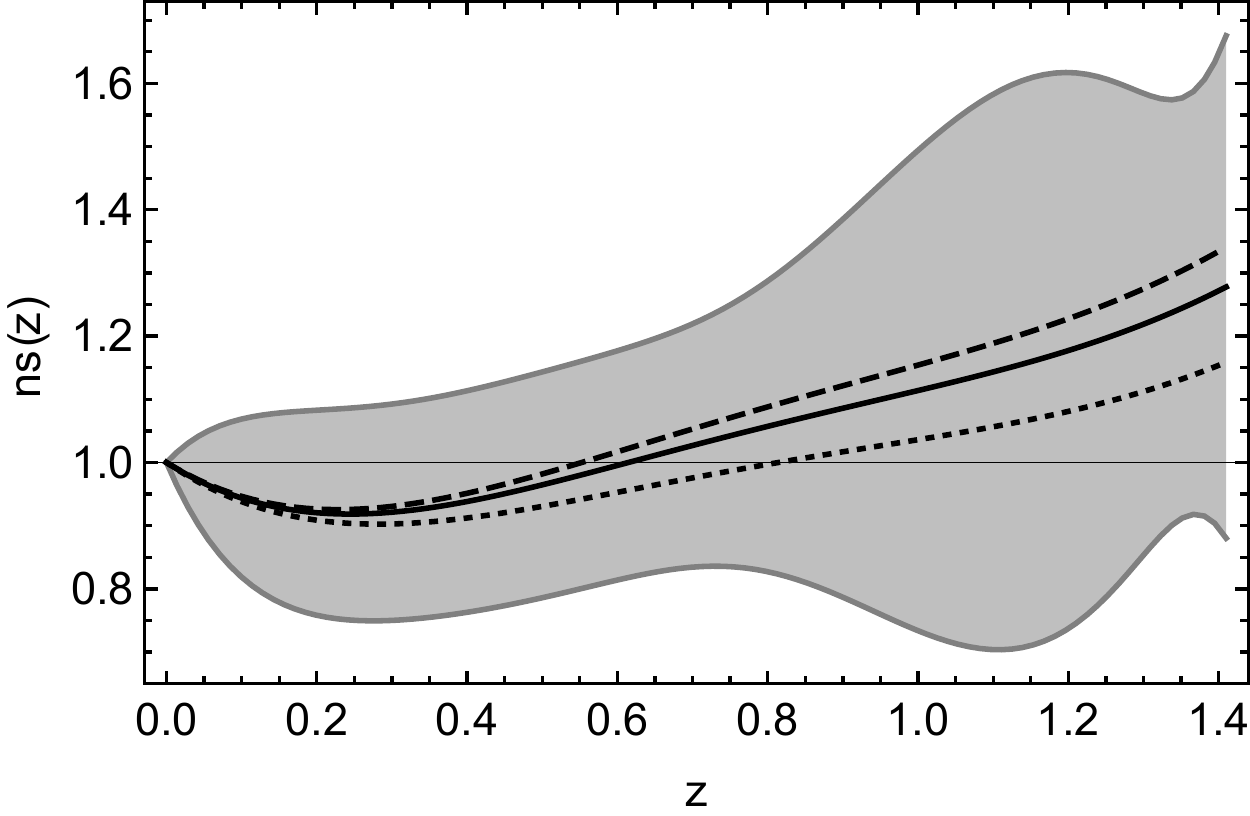}
\caption{
The four null tests considered in this work with 1$\sigma$ bands.
The flat $\Lambda$CDM model is falsified if the $Om$ diagnostics of Section~\ref{Om} do not give a constant value; a reference value of $\Omega_{m_{0}}=0.3$ is shown (solid line) to guide the eye. All the FLRW models are ruled out if the $Ok$ diagnostic of Section~\ref{oktest} is not compatible with a constant value (the solid line shows the reference value $\Omega_{k_{0}}=0$).
Finally, the (possibly curved) $\Lambda$CDM model is falsified if the $ns$ diagnostic of Section~\ref{NStest} is incompatible with a constant value of unity.
As these plots show, the latest cosmological data passes all the standard model null tests. For the $Om_{1}$ and $Ok$ tests we adopted the following values of the Hubble constant: the $H_{0,\rm{cmb}}$ value of equation~\eqref{cmb} (solid line and gray bands), $H_{0}= 70 \text{ Km }\text{s}^{-1} \text{Mpc}^{-1}$ (dashed line) and the $H_{0,\rm{loc}}$ value of equation~\eqref{riess} (dot-dashed line).
For the $ns$ test we adopt the values given in equation \eqref{s8va}: $\sigma_{8,0} = 0.804$ (black solid line), $\sigma_{8,0} = 0.832$ (black dashed line) and $\sigma_{8,0} = 0.747$ (black dotted line).
See Section~\ref{results} for more details.}
\label{fom}
\end{centering}
\end{figure*}

%%%%%%%%%%%%%%%%%%%%%%%%%%%%%%%%%%%%
\subsection{$Om$ diagnostic}\label{omr}
%%%%%%%%%%%%%%%%%%%%%%%%%%%%%%%%%%%%

Figure~\ref{fom} (left panels) show the reconstruction of the diagnostics $Om_{1}$ and $Om_{2}$.
The 1$\sigma$ gray bands are obtained using equation~\eqref{linear}.
The $Om_{1}$ test is, within its error, consistent with the reference value of about $0.3$.
Figure~\ref{fom} was obtained adopting the $H_{0,\rm{cmb}}$ value of equation~\eqref{cmb} (solid line and gray bands), $H_{0}= 70 \text{ Km }\text{s}^{-1} \text{Mpc}^{-1}$ (dashed line) and the $H_{0,\rm{loc}}$ value of equation~\eqref{riess} (dot-dashed line).

In order to evaluate the $Om_{2}$ test we need to specify the value of the nuisance parameter $M$, on which $D$ depends. As discussed in Section~\ref{additional}, $M$ is degenerate with $-5 \log_{10} H_{0}$. Therefore, we can fix $H_{0}$ to an arbitrary value such as $H_{0}= 70 \text{ Km }\text{s}^{-1} \text{Mpc}^{-1}$ and adopt the value of $M$ for which $Om_{2}$ is closest to a constant. If such a value of $M$ does not exist, then the flat $\Lambda$CDM model is falsified.
The lower left panel of Figure~\ref{fom} shows the $Om_{2}$ diagnostic for the value $M=0$.
Also this test is passed.
It is worth stressing that this test (and also the $Ok$ test discussed in the next section) will be more constraining when the comoving distance $D$ will be reconstructed using data which does not need unconstrained nuisance parameters such as $M$.
Future BAO data could be useful in this respect.

The $Om_{1}$ test (and also the $Ok$ test) diverges at $z=0$ as it involves a $0/0$ limit.
Because of the noise in the data, this limit is not well behaved and it does not follow the theoretical behavior.
Nonetheless, this is not a problem as also the errors diverge so that this singular behavior is correctly taken into account and does not bias the result.

%%%%%%%%%%%%%%%%%%%%%%%%%%%%%%%%%%%%
\subsection{$Ok$ diagnostic}\label{okr}
%%%%%%%%%%%%%%%%%%%%%%%%%%%%%%%%%%%%

In Fig.~\ref{fom} (top right panel) we show the results of the reconstruction of the $Ok$ diagnostic of Eq.~\eqref{ok}.
As with the $Om_{2}$ test of the previous section, one has to specify the value of the nuisance parameter $M$.
Moreover, in this case, the Hubble constant $H_{0}$ appears directly in the equation~\eqref{ok}.
In the plot we show the $Ok$ diagnostic that is obtained when adopting the $H_{0,\rm{cmb}}$ value of equation~\eqref{cmb} (solid line and gray bands), $H_{0}= 70 \text{ Km }\text{s}^{-1} \text{Mpc}^{-1}$ (dashed line) and the $H_{0,\rm{loc}}$ value of equation~\eqref{riess} (dot-dashed line).
It is clear that the reconstruction is consistent with $\Omega_{k_{0}} = 0$. 
The results found are consistent with those presented in~\cite{Sapone:2014nna}.

%%%%%%%%%%%%%%%%%%%%%%%%%%%%%%%%%%%%
\subsection{$ns$ diagnostic}\label{nsr}
%%%%%%%%%%%%%%%%%%%%%%%%%%%%%%%%%%%%

In Figure~\ref{fom} (lower right panel) we show the results relative to the $ns$ null test of equation~\eqref{ns}.
Present-day data again confirm within $1\sigma$ errors that the $\Lambda$CDM model is viable.
In this case, for consistency~\cite{Nesseris:2014mfa}, $H(0)$ is $H(z=0)$, where the latter is the Hubble function reconstructed from cosmic chronometer data.
We could also use the $H_0$ values used for the other tests; however, the effect is an overall shift of the entire curve as $H(0)$ enters as a multiplicative factor.
The $\sigma_{8,0}$ adopted (black solid line) is from the SDSS-III BOSS survey but we also plotted the $ns$ test for the $\sigma_{8,0}$ values from KiDS, see Section~\ref{additional}.

Next-generation surveys are expected to be able to accurately reconstruct this test: for example, $H(z)$ data will be obtained via galaxy clustering and $f\sigma_{8}(z)$ data via independent weak lensing observations.

%%%%%%%%%%%%%%%%%%%%%%%%%%%%%%%%%%%%
%%%%%%%%%%%%%%%%%%%%%%%%%%%%%%%%%%%%
\section{Conclusions} \label{conclusions}

In this paper, using $H(z)$, Supernova Ia and $f\sigma_8$ data we have reconstructed four null-tests, that can be used to probe deviations from either $\Lambda$CDM or the assumption of homogeneity and isotropy in the universe. The reconstruction has been performed by fitting the data with the linear model formalism which provides an exact statistical description of the reconstructed functions together with their derivatives and integrals.
We find that all the four tests are in agreement with the standard cosmological model and no interesting deviations were found.
This also implies that we have not found any tension on the data. 
However, a special attention goes to the value of $H_0$ used. As shown, the $Om_1$ test is the most sensitive to $H_0$ due to its direct dependence; from Fig.~\ref{fom} we understand that lower values of $H_0$ favor a larger value of $Om_1$. The other tests are less affected to the value of~$H_0$.

Current data give results that are far from the cosmic variance limit and future data will revolutionize the usefulness of the null tests: hundreds of thousands of supernovas Ia, hundreds of millions of galaxy spectra and even more shape measurements, 21-cm data to high redshifts will enable us to accurately use the null tests to assess the viability of the standard model of cosmology. In order to be ready for these large datasets it is important to develop and explore alternative methodologies to reconstruct the relevant cosmological functions.

\begin{acknowledgments}
It is a pleasure to thank Michele Moresco for clarifications regarding cosmic chronometer data.
VM thanks CNPq and FAPES for partial financial support. 
DS acknowledges financial support from the Fondecyt project number 11140496.
\end{acknowledgments}

%%%%%%%%%%%%%%%%%%%%%%%%%%%%%%%%%%%%
%%%%%%%%%%%%%%%%%%%%%%%%%%%%%%%%%%%%
%\bibliographystyle{plain}
\bibliographystyle{utphys}
\bibliography{references}
 %%%%%%%%%%%%%%%%%%%%%%%%%%%%%%%%%%%%
%%%%%%%%%%%%%%%%%%%%%%%%%%%%%%%%%%%%

%%%%%%%%%%%%%%%%%%%%%%%%%%%%%%%%%%%%
%%%%%%%%%%%%%%%%%%%%%%%%%%%%%%%%%%%%

\end{document}